\documentclass[aps,prl,twocolumn,groupedaddress]{revtex4}
\usepackage{graphicx}

\begin{document}
\title{Dynamic transition of supercritical hydrogen: defining the boundary between interior and atmosphere in gas giants}
\author{K. Trachenko$^{1}$}
\author{V. V. Brazhkin$^{2}$}
\author{D. Bolmatov$^{1}$}
\address{$^1$ South East Physics Network and School of Physics and Astronomy, Queen Mary University of London, Mile End Road, London, E1 4NS, UK}
\address{$^2$ Institute for High Pressure Physics, RAS, 142190, Moscow, Russia}

\begin{abstract}
Understanding physics of gas giants requires the knowledge about the behavior of hydrogen at extreme pressures and temperatures. Molecular hydrogen in these planets is supercritical, and has been considered as a physically homogeneous state where no differences can be made between a liquid and a gas and where all properties undergo no marked or distinct changes with pressure and temperature, the picture believed to hold below the dissociation and metallization transition. Here, we show that in Jupiter and Saturn, supercritical molecular hydrogen undergoes a dynamic transition around 10 GPa and 3000 K from the ``rigid'' liquid state to the ``non-rigid'' gas-like fluid state at the Frenkel line recently proposed, with accompanying qualitative changes of all major physical properties. The consequences of this finding are discussed, including a physically justified way to demarcate the interior and the atmosphere in gas giants.
\end{abstract}

\maketitle

Most abundant element in the Universe, hydrogen has been the subject of the cross-disciplinary research including the areas of condensed matter, astronomy and astrophysics. Despite its seeming simplicity, hydrogen continues to surprise by its rich and non-trivial behavior, particularly at high pressure and temperature at which hydrogen exists in gas giants such as Jupiter and Saturn as well as in hot gaseous exoplanets and brown dwarfs \cite{guillot,ceperley}. This is witnessed by advanced modeling and cutting-edge experimental compression techniques as well as space probes deployed recently to understand main physical mechanisms at operation in gas giants \cite{guillot,ceperley,mod1,mod11,mod12,mod2,mod3,mod4,mod5,mod6,mod7,melting,mod8,mod9,mod10}.

In gas giants such as Jupiter, Saturn, exoplanets and brown dwarfs, molecular hydrogen is supercritical (critical temperature and critical pressure of molecular hydrogen are about 33 K and 1.3 MPa), and this fact has been viewed according to the existing view of the homogeneity of supercritical state in terms of physical properties: moving along any path on a pressure and temperature phase diagram above the critical point does not involve marked changes of properties which vary only gradually and in a featureless way \cite{su1}. In gas giants, this has been considered \cite{guillot1} to be the case up to high 100-200 GPa pressures where hydrogen fluid dissociates and metallizes \cite{ceperley}, and has served as an important starting point of advance modelling and theoretical techniques \cite{guillot,mod1,mod11,mod12,mod2,mod3,mod4,mod5,mod6,mod7,mod8,mod9,mod10,guillot1,french}.

Several interesting open questions exist in the area \cite{guillot,ceperley,mod1,mod11,mod12,mod2,mod3,mod4,mod5,mod6,mod7,mod8,mod9,mod10,guillot1}, and one basic question is related to the supercritical nature of molecular hydrogen, namely whether and how a boundary between the planet's interior and exterior (atmosphere) can be defined? Unlike in terrestrial-type planets such as Earth and Venus where the boundary is clear, the boundary in gas giants is considered as conditional only because the supercritical state has been viewed as physically homogeneous and smooth. For practical purposes, the boundary between the interior and the atmosphere is conditionally taken at the pressure of the Earth atmosphere of 1 bar \cite{guillot,guillot1}. This gives the radius of about 70,000 km for Jupiter and 57,000 km for Saturn, in approximate agreement with measured optical sizes.

Recently, we have proposed \cite{phystoday,pre,natcom,velcor,jchem} that the supercritical state is not physically homogeneous, but exists in two states with qualitatively distinct physical properties. The two states are separated by the dynamic transition at the Frenkel line on the phase diagram. This raises an important question. Depending on the ($P$,$T$) conditions in a gas giant, molecular hydrogen can be either always above the line or it can cross the line. The last scenario is most intriguing because it means that as we go inside the gas giant interior, the dynamic transition at the Frenkel line takes place, implying that supercritical molecular hydrogen exists in two physically distinct states. Here, we find that this scenario is realized in Jupiter and Saturn but not in larger exoplanets and brown dwarfs for which the data is available.

We start with a brief four-paragraph discussion of the dynamic transition at the Frenkel line. This is followed by discussing the relationship between the line and dynamic transition in gas giants.

In gases, particles move in almost straight lines until they change course due to collisions. In liquids, particle motion has two components: a solid-like, quasi-harmonic vibrational motion about equilibrium positions and diffusive ballistic jumps between neighboring equilibrium positions \cite{frenkel}. As the temperature increases or the pressure decreases, a particle spends less time vibrating and more time diffusing. Eventually, the solid-like oscillating component of motion disappears; all that remains is the ballistic-collisional motion. That disappearance, a qualitative change in particle dynamics, takes place at particular values of pressure and temperature; the collection of these points define the line on pressure and temperature phase diagram, the Frenkel line \cite{phystoday,pre}. Crossing the Frenkel line corresponds to the dynamic transition discussed below. Importantly, the Frenkel line exists at arbitrarily high pressure and temperature above the critical point (with a caveat that chemical and electronic changes such as ionization or metallization at higher pressure and temperature may result in a new dynamic line within a different phase), is universal for all fluids, and exists even in systems where the liquid-gas transition and the critical point are absent altogether \cite{pre,velcor}.

The first effect taking place at the Frenkel line is the loss of rigidity at all available frequencies. As predicted by Frenkel \cite{frenkel} and subsequently verified experimentally, liquids support shear stress and solid-like ``rigid'' transverse waves at frequency larger than $\frac{1}{\tau}$, where $\tau$ is liquid relaxation time, the average time between two consecutive atomic jumps in a liquid at one point in space. At the Frenkel line, $\tau$ approaches its minimal value of $\tau_{\rm D}$, where $\tau_{\rm D}$ is the shortest Debye vibration period of about 0.1 ps. At this point, the system simply cannot sustain rigidity at any frequency, and behaves like a gas. We therefore call the system below the Frenkel line ``rigid'' liquid and above the line ``non-rigid'' gas-like fluid \cite{pre}.

Importantly, most important system properties change qualitatively at the Frenkel line, as is evidenced by theory, simulations and supercritical experimental data \cite{phystoday,pre,velcor,natcom}. The speed of sound (see Figure 1), viscosity and thermal conductivity all decrease with increasing temperature below the Frenkel line as in liquids, but increase with temperature sufficiently above the line as in gases. The diffusion constant crosses over from exponential temperature dependence below the Frenkel line as in liquids, to power-law dependence above the line as in gases. Crossing the Frenkel line also results in the disappearance of fast sound and roton minima, both characteristic features of liquids \cite{pre}. We have also shown that heat capacity undergoes a crossover at the Frenkel line \cite{natcom}, coinciding with the disappearance of oscillations in velocity-velocity correlation function \cite{velcor}.

More recently, we have demonstrated that the structure of the supercritical matter undergoes a crossover at the Frenkel line, an unanticipated finding in view of the perceived structural homogeneity of the supercritical state \cite{jchem}. The Frenkel line demarcates liquid-like configurations with structural correlations in the medium range and gas-like configurations where these correlations are absent.

We now discuss the location of the Frenkel line in hydrogen at pressure and temperature conditions existing in giant gas planets, exoplanets and Brown dwarfs. We use two criteria and data sets to plot the Frenkel line. First, we use the temperature dependence of the isobaric speed of sound measured in supercritical H$_2$ from the NIST (National Institute of Standards and Technology) database (see http://webbook.nist.gov/chemistry/fluid). In Figure 1, we observe temperature minima, $T_{\rm min}$, of the speed of sound at different pressures $P$. As discussed earlier, the minima correspond to the dynamic transition at the Frenkel line \cite{phystoday,pre,velcor}. Indeed, the speed of sound in condensed phases such as solids and ``rigid'' liquids is primarily governed by interactions between atoms and elastic moduli, and is known to decrease with temperature. On the other hand, the speed of sound in the ballistic non-rigid gas-like regime is simply given by thermal velocity of particles increasing with temperature without bound. Therefore, sound velocity has a minimum close to the region where particle dynamics changes the character from combined oscillatory and ballistic to purely ballistic, as seen in Figure 1. We plot $P$ and $T_{\rm min}$ in Figure 2 as open blue diamonds.

\begin{figure}
\begin{center}
{\scalebox{0.42}{\includegraphics{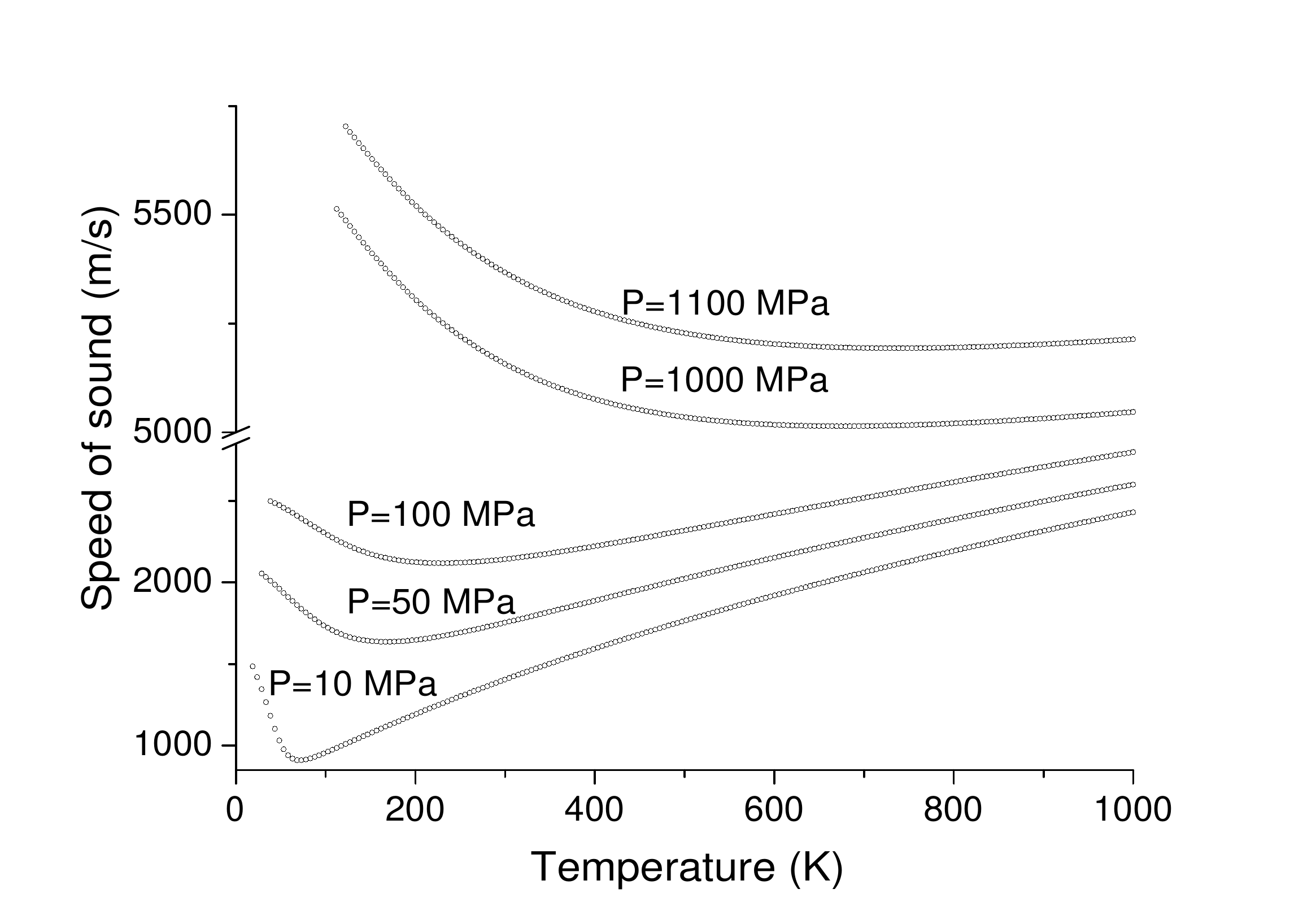}}}
\end{center}
\caption{Representative dependencies of the speed of sound in supercritical H$_2$ as a function of temperature at different pressures. The data are from the NIST database.}
\label{2}
\end{figure}

\begin{figure*}
\begin{center}
{\scalebox{0.7}{\includegraphics{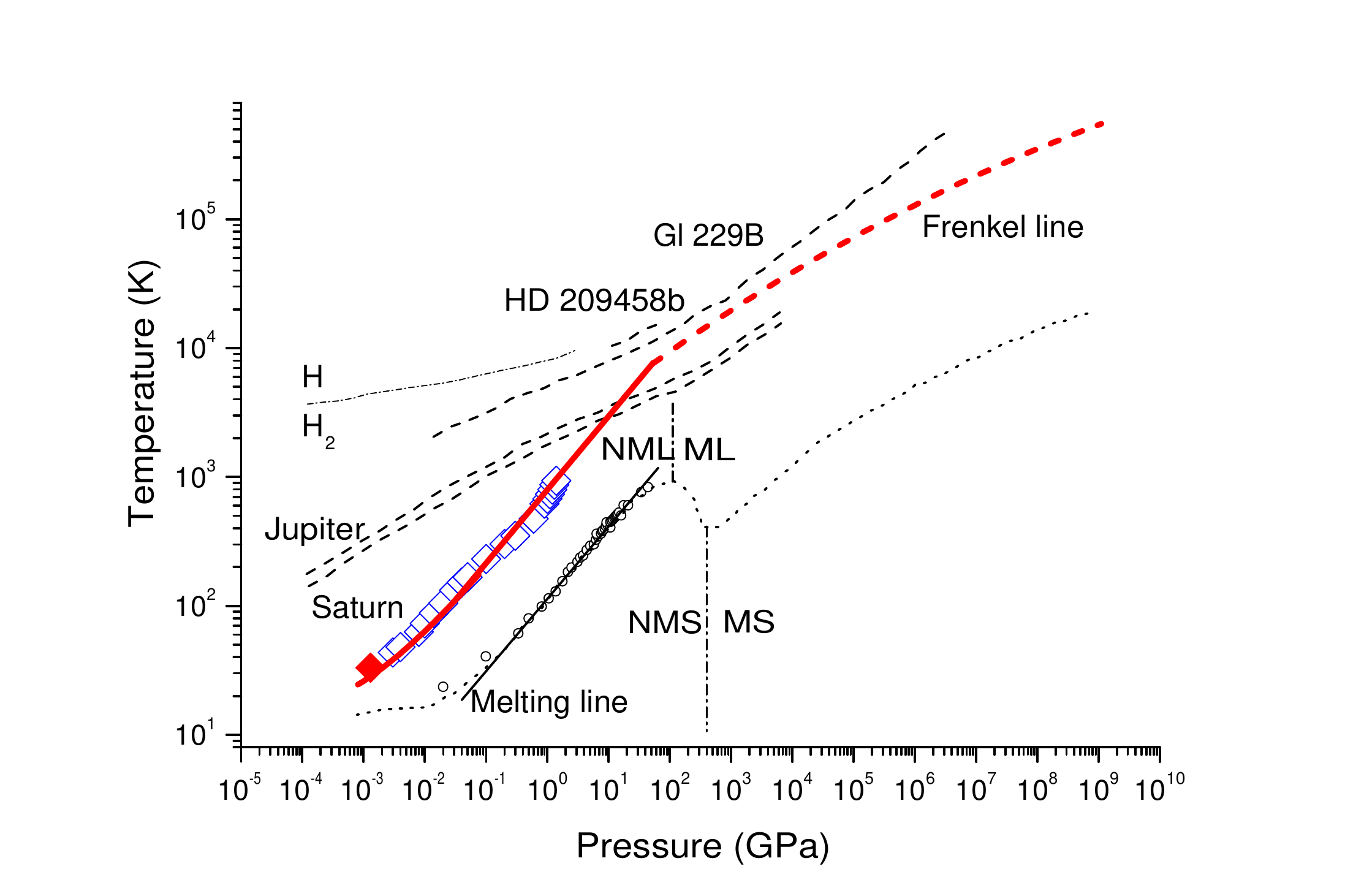}}}
\end{center}
\caption{The Frenkel line and adiabats in gas giants. Black dashed lines show pressure and temperature adiabats inside Jupiter, Saturn, exoplanet HD 209458b and brown dwarf G1 229B \protect\cite{ceperley}. Empty black bullets show the melting line of H$_2$ \protect\cite{melting}, with the straight line showing the linear range. Dotted line shows the melting line in a wider pressure and temperature range \protect\cite{ceperley}. Open blue diamonds correspond to pressure and temperature at the minima of the speed of sound from the NIST database shown in Figure 1, and solid red diamond shows the critical point of H$_2$. The thick red line shows the Frenkel line starting just below the critical point and running parallel to the melting line in the range 0.5--50 GPa and through the minima of the speed of sound. The dashed red line shows a tentative extrapolation of the Frenkel line in the high pressure range based on existing melting data. Dash-dotted line at low pressure separates molecular hydrogen H$_2$ and dissociated hydrogen H \protect\cite{ceperley}. Two vertical dash-dotted lines approximately separate non-metallic solid (NMS) and metallic solid (MS) at low temperature, and non-metallic liquid (NML) and metallic liquid (ML) at high temperature \protect\cite{ceperley}.}
\label{3}
\end{figure*}

We note that the minima in Figure 1 appear less pronounced at high pressure. This is due to the absence of NIST data above 1000 K and associated apparent reduction of the minima depth at high pressure (see Figure 1) rather than a change in physical behaviour. As discussed in detail \cite{phystoday,pre,velcor,yura}, the Frenkel line and the associated minima in the speed of sound extend to arbitrarily high temperature and pressure above the critical point. This takes place with the caveat mentioned above: if chemical and electronic changes are triggered at high pressure and temperature and a new phase is created (e.g., metallization \cite{ceperley}), a new dynamic line may emerge in the new state.

The NIST database does not extend to temperatures and pressures high enough to match those in gas giants, and therefore we use the second criterion: the Frenkel line starts slightly below the critical point and at high pressure is parallel to the melting line in the log-log plot \cite{pre,velcor}. The parallelism follows from the well-known scaling argument: at high pressures starting from GPa, the intermolecular interaction is reduced to its repulsive part only, whereas the cohesive attracting part no longer affects interactions (at low pressure, the parallelism between the two lines holds only approximately because the interactions are not well approximated by simple repulsive laws, see below). In a sufficiently wide pressure range, the repulsive part can be well approximated by several empirical interatomic potentials such as the Buckingham-type functions or Lennard-Jones potentials with inverse power-law leading terms at short distances, $U\propto\frac{1}{r^n}$ (see, e.g., \cite{belon}). For the inverse-power law, a well-known scaling of pressure and temperature exists: system properties depend only on the combination of $TP^\gamma$, where $\gamma$ is uniquely related to $n$. Consequently, $TP^\gamma=$const on all ($P$,$T$) lines where the dynamics of particles changes qualitatively, as it does on both the melting line and the Frenkel line. This implies that the Frenkel and melting lines are parallel to each other in the double-logarithmic plot. This is directly confirmed by the recent molecular dynamics simulations where the calculated Frenkel line is found to be parallel to the melting line for a number of supercritical fluids significantly above the critical point \cite{yura}. We therefore draw the Frenkel line parallel to the melting line (which is straight in approximately 0.5--50 GPa range) and ending just below the critical point. Importantly, in Figure 2 we observe that the Frenkel line constructed parallel to the melting line up to about 50 GPa lies closely to ($P$,$T_{\rm min}$) points defined from the minima of the speed of sound, serving as a self-consistency check in our construction.

We note that extending the Frenkel line to ultra-high pressures above 100 GPa can be done approximately only due to challenges and uncertainties of locating the melting line and nonmetal-metal transition in both liquid and solid phases of hydrogen \cite{ceperley}. For this reason, we draw the Frenkel line above 50 GPa as a dashed line that is approximately parallel to the melting line at ultrahigh pressures where hydrogen is monatomic and metallic.

In Figure 2 we observe that the Frenkel line crosses the adiabats of Saturn and Jupiter at approximately ($P_{\rm F}=10$ GPa, $T_{\rm F}=3000$ K) and ($P_{\rm F}=17$ GPa, $T_{\rm F}=3900$ K), respectively. Importantly, ($P_{\rm F}, T_{\rm F})$ are below pressure and temperature at which dissociation, metallization, ionization and other chemical and electronic transitions take place \cite{ceperley} (see also Figure 2). Our finding therefore implies that supercritical molecular hydrogen in Saturn and Jupiter exists in two physically distinct states: non-rigid gas-like fluid below ($P_{\rm F}$,$T_{\rm F}$) and rigid liquid above ($P_{\rm F}$,$T_{\rm F}$). The non-rigid gas-like fluid exists in the outer part of the planet, and is separated by the Frenkel line from the rigid liquid state located closer to the planet's centre up to the region where dissociation and metallization take place. Importantly, all main physical properties of supercritical molecular hydrogen in Jupiter and Saturn change qualitatively at the dynamic transition at the Frenkel line as discussed above. This constitutes the main finding of this paper.

Before proceeding further, we make two remarks. First, our finding concerns pressures well below 100--200 GPa at which metallization and dissociation of hydrogen take place \cite{ceperley} (see Figure 2). Our results are therefore not related to the transition to the atomic metallic phase, but to the molecular hydrogen in the supercritical state. Second, the Frenkel line inside a gas giant does not imply that a physical property (such as viscosity, thermal conductivity or the speed of sound) has a minimum as a function of the radial distance $R$ on crossing the line. Lets introduce $R_{\rm F}$ as the distance between the centre and the location of the Frenkel line and consider molecular hydrogen at points $R_{\rm F}-\Delta R$ and $R_{\rm F}+\Delta R$. Then, crossing the Frenkel line implies the following. At $R_{\rm F}-\Delta R$, supercritical molecular hydrogen is in the liquid-like state where the above physical properties decrease with temperature at constant pressure as discussed above. At $R_{\rm F}+\Delta R$, hydrogen is in the gas-like state where these properties increase with temperature at constant pressure.

Unlike in Jupiter and Saturn, the Frenkel line does not cross the adiabates of exoplanet HD 209458b and brown dwarf G1 229B that are larger and hotter. Consequently, supercritical molecular hydrogen in these gas giants is always in the non-rigid gas-like fluid state, the picture that also holds for even hotter objects such as the Sun.

The discovered dynamic transition has important consequences for understanding the physical processes in gas giants. Generally, the transition and associated property changes should be incorporated in theory and advanced planetary modeling where an active research is ongoing and proposals to resolve controversies related to most basic properties of gas giants are discussed \cite{guillot,ceperley,mod1,mod11,mod12,mod2,mod3,mod4,mod5,mod6,mod7,mod8,mod9,mod10,french}. In particular, the qualitative changes of diffusion, viscosity and thermal conductivity at the Frenkel line fundamentally affect flow, convection and heat transport processes, the processes that are at the centre of ongoing theory and modeling of gas giants \cite{mod1,mod2,mod3,mod7,mod10,mod11,mod12,guillot1}.

The dynamic transition at the Frenkel line can serve as a physically justifiable boundary between the interior and the exterior (or, if appropriate, the atmosphere) of gas giants containing supercritical matter. Indeed, the outer matter in planets such as Earth and Venus is below the critical point, and the boundary is clearly defined as the boundary between the gas on one hand and the liquid or solid on the other, with liquid-gas and liquid-solid first-order phase transitions separating these states and the character of atomic motion being distinctly different in all three states: pure oscillations in solids, pure ballistic motion in gases and mixed oscillation and diffusional motion in liquids. On the other hand, matter in gas giants is supercritical so that no first-order transition demarcates the phase boundaries, making the separation between the interior and exterior problematic from this perspective. In gas giants in the solar system, this demarcation is often done by specifying some low pressure conditionally taken to be equal to the pressure of the Earth's atmosphere of 1 bar \cite{guillot}. Convenient for the purposes of comparison, this definition is arbitrary from the physical point of view.

Here, we have shown that a physical boundary of supercritical hydrogen exists in gas giants at the Frenkel line at which the character of particle motion qualitatively changes, with the accompanied qualitative changes of particle dynamics and associated dramatic changes of all major physical properties from gas-like to liquid-like. This is exactly the same as what takes place at the liquid-gas planet-atmosphere boundary in smaller planets such as Earth or Venus except without a first-order phase transition. Among the properties changing qualitatively at the Frenkel line, rigidity, the stability against solid-like shear distortions, is particularly meaningful here. Indeed, on planets such as Earth and Venus, rigidity changes on the surface, and separates planet's interior from the atmosphere because the solid crust and the liquid ocean possess rigidity (in solids rigidity is static and in liquids it is dynamic at frequencies $\omega>\frac{1}{\tau}$ as discussed above) whereas the gas state does not. In gas giants where the matter is supercritical, it is exactly rigidity that exists below but not above the Frenkel line as discussed above.

For the above reasons, we propose that the Frenkel line serves as a physically justified boundary between the interior and exterior (or atmosphere) of gas giants. Particularly relevant to this proposal is the recent evidence for the structural crossover at the Frenkel line that separates the liquid-like configuration with structural correlations in the medium range and gas-like configuration where these correlations are absent \cite{jchem}.

Using the relationship between pressure and density \cite{guillot1}, values of $P_{\rm F}$ above and the hydrostatic relationship $P_{\rm F}=g\rho H_{\rm F}$, where $H_{\rm F}$ is the height of the Frenkel line boundary below the current 1 bar ``surface'', we approximately estimate $H_{\rm F}$ to be ($4000\pm 1000$) km and $(6000\pm 1000$) km in Jupiter and Saturn, respectively, corresponding to 66,000 km from the centre in Jupiter (94\% of the currently used radius) and 51,000 km in Saturn (89\% of the current radius).

\acknowledgments
K. Trachenko is grateful to EPSRC, and V. V. Brazhkin to RFBR for financial support. We are grateful to I. Polichtchouk, J. Cho, R. Nelson and J. Schneider for discussions.

\end{document}